\documentclass[fleqn,usenatbib]{mnras}

\usepackage{newtxtext,newtxmath}

\usepackage[T1]{fontenc}

\DeclareRobustCommand{\VAN}[3]{#2}
\let\VANthebibliography\thebibliography
\def\thebibliography{\DeclareRobustCommand{\VAN}[3]{##3}\VANthebibliography}

\defcitealias{Maragkoudakis2020}{MPR20}
\defcitealias{Maragkoudakis2023}{MPR23}

\usepackage{graphicx}	
\usepackage{amsmath}	
\usepackage{subcaption}
\usepackage{hyperref}
\usepackage{ulem}

\usepackage{xcolor}

\newcommand{\Nc}{N$_{\mathrm{C}}$}
\newcommand{\AvgNc}{$\langle\mathrm{N_{C}}\rangle$}

\title[PAH Size Tracers]{Polycyclic Aromatic Hydrocarbon Size Tracers}

\author[A. Maragkoudakis et al.]
{A. Maragkoudakis$^{1,2}$\thanks{E-mail: alexandros.maragkoudakis@nasa.gov}, E. Peeters$^{3,4,5}$, A. Ricca$^{1,5}$, C. Boersma$^{1}$\\
    $^{1}$NASA Ames Research Center, MS 245-6, Moffett Field, CA 94035-1000, USA\\
    $^{2}$Oak Ridge Associated Universities, Oak Ridge, TN, USA\\
	$^{3}$Department of Physics and Astronomy, University of Western Ontario, London, ON, N6A 3K7, Canada\\
	$^{4}$Centre for Planetary Science and Exploration, University of Western Ontario, London, Ontario N6A 5B7, Canada \\
	$^{5}$Carl Sagan Center, SETI Institute, 339 Bernardo Ave., Mountain View, CA 94043, USA}

\date{Accepted XXX. Received YYY; in original form ZZZ}

\pubyear{2023}

\begin{document}
\label{firstpage}
\pagerange{\pageref{firstpage}--\pageref{lastpage}}
\maketitle

\begin{abstract}

We examine the dependence of polycyclic aromatic hydrocarbon (PAH) band intensity ratios as a function of the average number of carbon atoms and assess their effectiveness as tracers for PAH size, utilising the data, models, and tools provided by the NASA Ames PAH Infrared Spectroscopic Database. To achieve this, we used spectra from mixtures of PAHs of different ionisation fractions, following a size distribution. Our work, congruent with earlier findings, shows that band ratios that include the 3.3~\micron{} PAH band provide the best PAH size tracers for small-to-intermediate sized PAHs. In addition, we find that band ratios that include the sum of the 15-20~\micron{} PAH features (I$_{\Sigma_{15-20}}$) and the 6.2 or 7.7~\micron{} bands also serve as good tracers for PAH size in the case of small-to-intermediate sized PAHs, for objects under a similar PAH size distribution as with the presented models. For different PAH size distributions, the application of a scaling factor to the I$_{6.2}$/I$_{\Sigma_{15-20}}$ ratio can provide estimates for the size of the small-to-intermediate PAH population within sources. Employment of the I$_{6.2}$/I$_{\Sigma_{15-20}}$ and I$_{7.7}$/I$_{\Sigma_{15-20}}$ ratios can be of particular interest for \textit{JWST} observations limited only to $\sim$5-28~\micron{} MIRI(-MRS) coverage. 
\end{abstract}




\section{Introduction}

Prominent emission features attributed to a family of organic aromatic molecules known as polycyclic aromatic hydrocarbons \citep[PAHs;][]{Leger1984, Allamandola1985} are observed in the 3-20~\micron{} spectra of a wide variety of astrophysical sources, including planetary and reflection nebulae, disks around young stars, the general interstellar medium, star-forming regions, and galaxies. PAHs are an energetically important component of interstellar dust, responsible for ${\sim}5-20\%$ of the total infrared (IR) emission in galaxies \citep[][]{Smith07b}. The emission is produced when PAHs become vibrationally excited after the absorption of UV/Vis photons and subsequently relax through radiating in the IR, producing distinct and prominent bands at 3.3, 6.2, 7.7, 8.6, 11.2, and 12.7~\micron{}, as well as weaker (satellite) features \citep[e.g.,][]{Tielens2008}.

PAH emission has been broadly utilised to characterise the local and larger-scale environments they inhabit \citep[e.g.,][]{Hony2001, Peeters2002, Smith07b, Galliano2008b, Gordon2008, Sandstrom2012, Pilleri2012, Peeters2017, Boersma2018, Maragkoudakis2018b, Zang2022}. This is possible because PAHs are highly susceptible to changing physical conditions in the regions they inhabit, such as metallicity, radiation field, gas temperature, and electron density. All these factors have a direct impact on their abundance, charge state, and size distribution. Consequently, determining the make-up of the astronomical PAH population and its spectral response to changing conditions has become an essential part for characterising astronomical environments.

Empirical calibrations of PAH band strength ratios are frequently used to estimate the PAH ionisation fraction and size distribution. These calibrations rely on assumptions, for example, regarding the PAH size distribution, the stochastic heating of PAHs, the absorption efficiencies, etc. \citep[][]{Schutte1993, BregmanTemi2005, Draine:07, Galliano2008b, Mori2012, Stock2016, Draine2021}.

Alternatively, the PAH charge state and size distribution in astrophysical sources can be inferred by employing a database-fitting approach \citep[e.g.,][]{Cami2010, Andrews2015, Boersma2018, Maragkoudakis2022}, and empirical calibrations of PAH properties can be obtained via construction of PAH populations from the ground up \citep[][]{Ricca2012, Croiset2016, Maragkoudakis2020, Knight2022, Maragkoudakis2023}.
The NASA Ames PAH Infrared Spectroscopic Database\footnote{\href{www.astrochemistry.org/pahdb/}{www.astrochemistry.org/pahdb/}} \citep[PAHdb hereafter;][]{Bauschlicher2010, Boersma2014b, Bauschlicher2018, Mattioda2020} is well-suited for this purpose, as it holds the world's foremost collection of laboratory-measured and quantum-chemically computed PAH spectra and offer a suite of modelling and (software) analysis tools. Its library of computed spectra, holds spectra of PAHs with various structures, charge states, sizes, and compositions.

In this work, we utilise PAHdb data, models, and tools to examine a range of PAH band intensity ratios to assess their prospective use as tracers for PAH size, offering optimal and alternative PAH size proxies. The importance of alternative tracers is especially relevant for observations with limited wavelength coverage. The \textit{James Webb Space Telescope} (\textit{JWST}), for example, is currently delivering a wealth of spectro-photometric observations of PAH-rich sources \citep[e.g.,][]{Evans2022, U2022, Lai2022, Chastenet2022, Chastenet2023, Egorov2023, Dale2023, Sandstrom2023}, enabling the characterisation of the PAH population even in spatially resolved regions of extra-galactic sources, and requires two separate instruments and observing modes to cover the entire near-to-mid IR wavelength range. \textit{JWST} observations constrained to the 5-28~\micron{} region (MIRI-MRS) are missing one of the key features for PAH size characterisation, i.e., the 3.3~\micron{} PAH band. For cases where \textit{JWST} NIRSpec observations are not feasible, alternative PAH size tracers are needed.

The paper is organised as follows: Section~\ref{sec:method_analysis} presents an overview of the methodology used to generate PAH mixtures and derive their spectra. In Section~\ref{sec:results} we describe the results and examine the application of the models to observations. A summary and conclusions is given in Section~\ref{sec:conclusions}.

\section{Methodology And Analysis} \label{sec:method_analysis}

The overall methodology and analysis in this work is that followed in \citet*[][\citetalias{Maragkoudakis2023} hereafter]{Maragkoudakis2023}. Specifically, we use the set of PAH spectra defined in \citetalias{Maragkoudakis2023}, which in turn is based on the sample of individual PAH molecules defined by \citet*[][\citetalias{Maragkoudakis2020} hereafter]{Maragkoudakis2020}. These spectra were selected from version 3.20 of PAHdb's library of density functional theory (DFT) computed spectra. The \citetalias{Maragkoudakis2020} set include pairwise neutral and singly charged PAH cations and were selected to meet astronomically relevant criteria. That is, (i) have more than 20 carbon atoms (\Nc), which is assumed to be the minimum to be able to survive typical interstellar radiation fields \citep[e.g.,][]{Allamandola1989, Puget1989, Allain1996}; (ii) contain no nitrogen, oxygen, magnesium, or iron atoms, in order to examine only ``pure'' PAH populations; and (iii) have solo C-H bonds to ensure the inclusion of PAHs able to produce the strong astronomical 11.2~\micron{} emission feature, which is frequently employed to quantify PAH charge and size through PAH intensity ratios. The range of \Nc{} of the individual PAHs is between 22-216 carbon atoms. For the generation of the emission spectra we consider PAHs exposed to the interstellar radiation field from \citet*{Mathis1983}, where the entire emission cascade is taken into account, and the emission bands are convolved using Lorentzian line profiles with a full-width-at-half-maximum of 15~cm$^{\rm -1}$.

The methodology used to synthesise PAH spectra of varying \Nc{} and ionisation fractions is that described in \citetalias{Maragkoudakis2023}, and is briefly summarised here. The individual PAH spectra from \citetalias{Maragkoudakis2020} are first separated into bins based on their size, in terms of \Nc. Subsequently, for each bin the average spectrum is calculated from the combination of the individual spectra within that bin, further weighted with the power-law size distribution from \cite{Schutte1993}.  The average \Nc{} (\AvgNc{}) is also determined for each bin. Then, the total spectrum is calculated by summing the weighted averaged spectrum of each bin. In total, two data sets are created by successively subtracting the averaged weighted spectrum of each bin from the total spectrum by: (i) starting from the smallest size bin and one-by-one subtracting the adjacent bin when moving to larger bins, and (ii) starting from the largest size bin and one-by-one subtracting the adjacent bin when moving to smaller bins. These data sets represent two scenarios for PAH evolution, namely: (1) small PAHs are being removed (or equivalently large PAHs are being formed), and (2) large PAHs are being removed (or equivalently small PAHs are being formed). A drop in the intensity among the two scenarios at $\sim$110 \AvgNc{} is evident due to the removal of the first bin of small PAHs which has the highest weight (see \citetalias{Maragkoudakis2023}). In addition to the purely neutral or cationic PAHs, spectra are synthesised from combining a neutral spectrum with that of its cation counterpart with relative neutral--cation contributions of: (i) 75-25 percent (N75C25), (ii) 50-50 percent (N50C50), and (iii) 25-75 percent (N25C75). As a result, PAH spectra of different \AvgNc{} and ionisation fractions are generated. Here we present results for the N50C50 case. The remainder of ionisation fractions are presented in Appendix~\ref{app:var_fi}.  

The intensities of the prominent computed PAH emission bands at 3.3, 6.2, 7.7, 8.6, and 11.2~\micron{} are measured following \citetalias{Maragkoudakis2020}, i.e., as the sum of fluxes in predefined wavebands at a fixed wavelength interval (see \citetalias{Maragkoudakis2020}, their Table~1, for details). In addition, the 15-20~\micron{} intensity (I$_{\Sigma_{15-20}}$) is measured, which encompasses prominent astronomical features at 16.4 and 17.4~\micron{}, weaker bands at 15.8 and 18.9~\micron{}, and a broad component centred at ${\sim}17$~\micron{} \citep[][]{VanKerckhoven:00, Smith07b, Boersma2010, Peeters2012, Shannon2015}.

\section{Results And Discussion} \label{sec:results}

Here, we examine and discuss the PAH band intensity ratios that are considered \emph{optimal} PAH size tracers, i.e., that have the strongest and most well-defined dependence on \AvgNc, as well as \emph{alternative} PAH size tracers that can be employed instead of the optimal tracers, depending on the observational wavelength coverage.

\subsection{Optimal PAH size tracers} \label{subsec:optimal_tracers}

\begin{figure}
    \centering
    \includegraphics[scale=0.52]{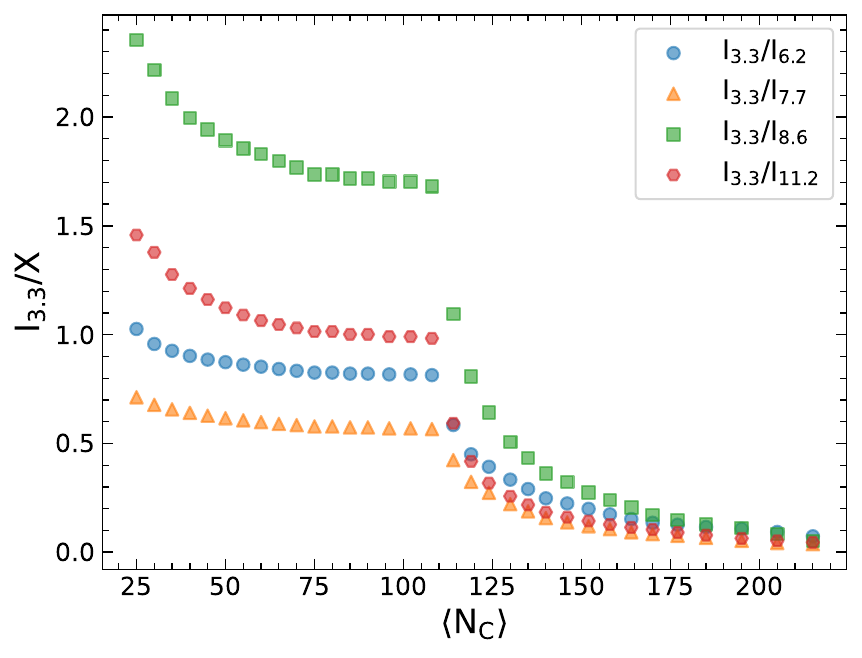}
    \caption{The dependence of the $I_{3.3}$ over the $I_{6.2}$ (blue), $I_{7.7}$ (orange), $I_{8.6}$ (green), and $I_{11.2}$ (red; presented in \citetalias{Maragkoudakis2023}) PAH band intensities on \AvgNc{}. The discontinuities at $\sim$110 \AvgNc{} are due to the different data sets used corresponding to different cases of PAH processing (see~Section~\ref{sec:method_analysis} and \citetalias{Maragkoudakis2023} for details.)}
    \label{fig:PAH33_X}
\end{figure}

The $I_{3.3}$/$I_{11.2}$ PAH band intensity ratio is widely considered the most robust PAH size tracer due to the large difference in attained energy between the 3.3 and 11.2~\micron{} bands, and because both bands are affected similarly by charge, and their emission originates from C-H modes \citep[e.g.,][]{Allamandola1989, Jourdain1990, Schutte1993, Mori2012, Ricca2012, Croiset2016, Maragkoudakis2020, Knight2021,  Maragkoudakis2022}. In addition, \citetalias{Maragkoudakis2020} showed that, in the case of single PAH molecules, the correlation between the $I_{3.3}$/$I_{6.2}$, $I_{3.3}$/$I_{7.7}$, $I_{3.3}$/$I_{8.6}$, and $I_{3.3}$/$I_{11.2}$ PAH band intensity ratios scale with \Nc{} across two orders of magnitude for both neutral and cationic PAHs, with the $I_{3.3}$/$I_{11.2}$ ratio having the strongest correlation. In this work we show that these correlations hold for PAH mixtures following a size distribution as well, thus providing a more accurate representation of the PAH population size in astronomical sources. Figure~\ref{fig:PAH33_X} presents these intensity ratios as a function of \AvgNc{}.

\begin{figure*}
    \centering
    \includegraphics[scale=0.4]{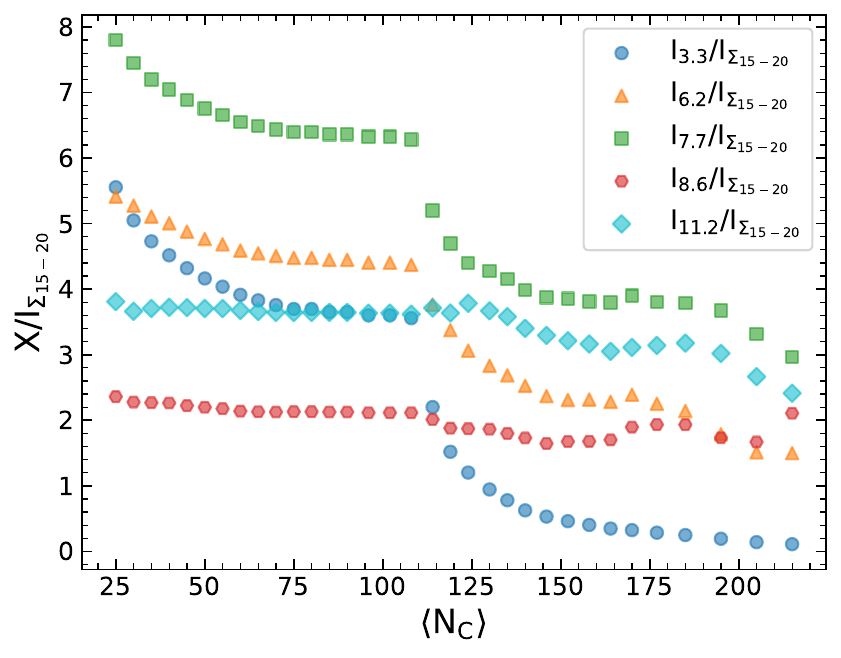}
    \includegraphics[scale=0.4]{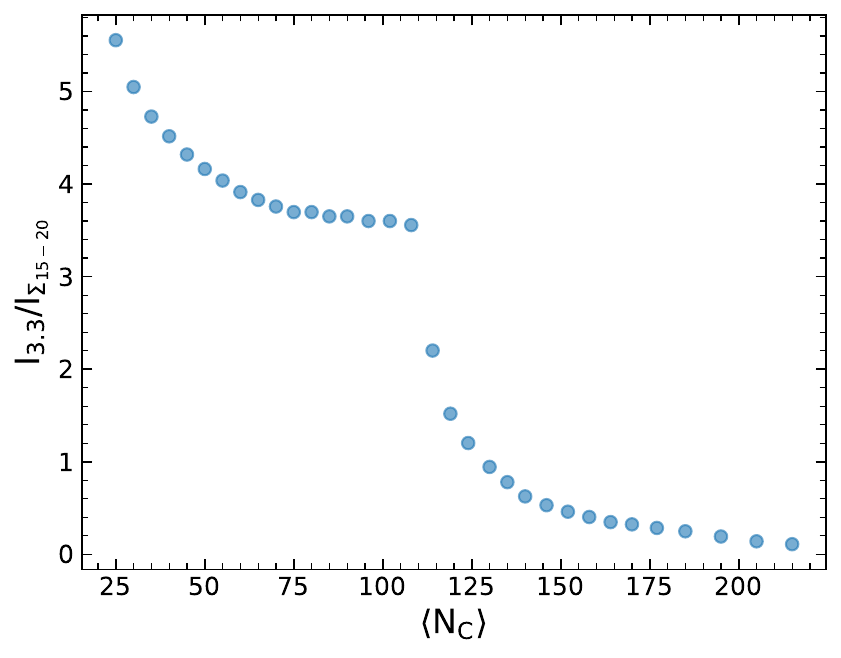}
    \includegraphics[scale=0.4]{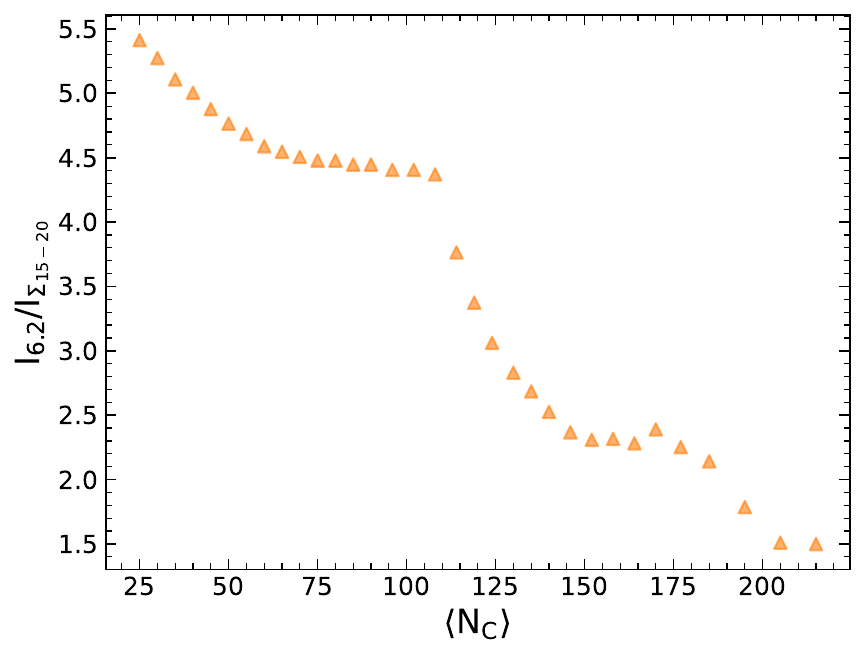} \\
    \includegraphics[scale=0.4]{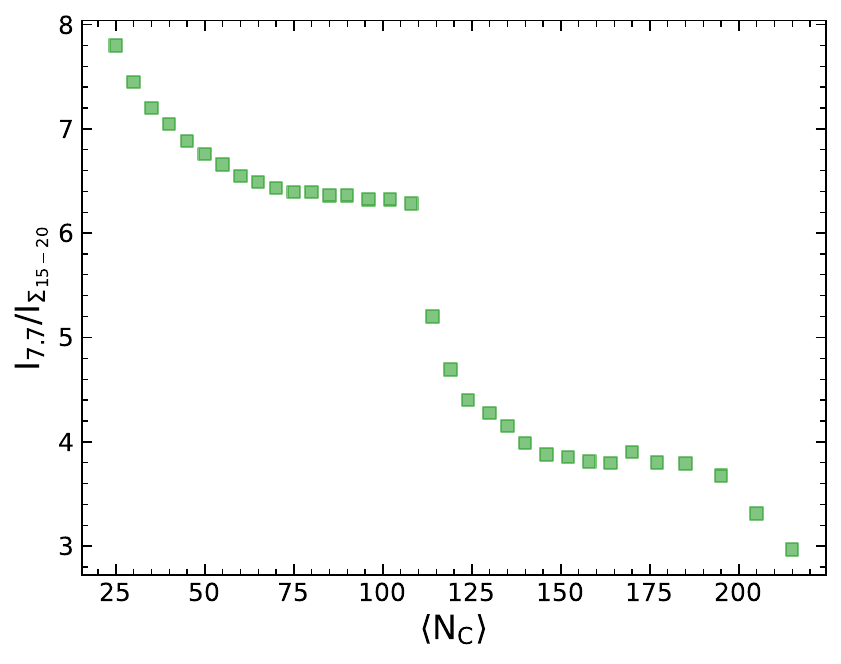} 
    \includegraphics[scale=0.4]{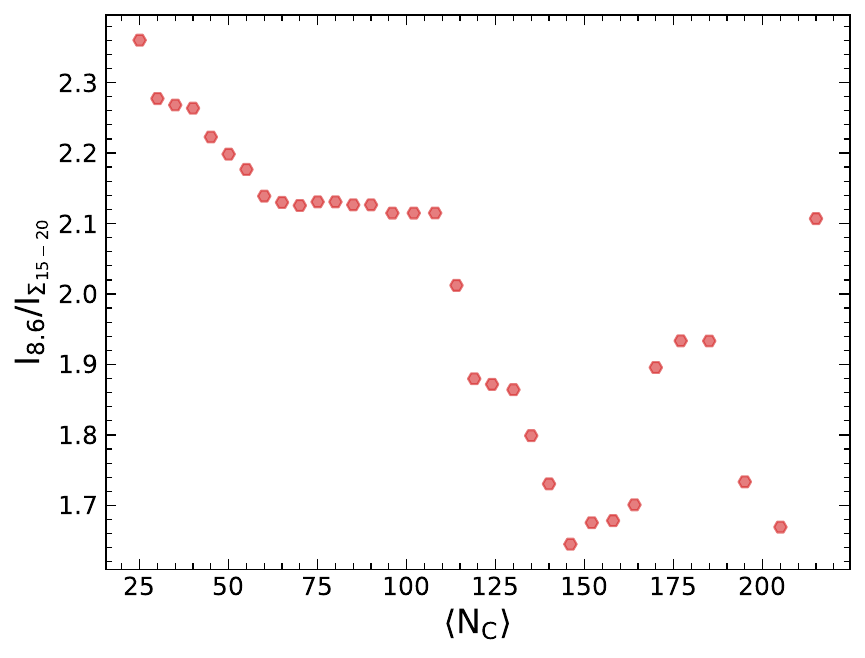} 
    \includegraphics[scale=0.4]{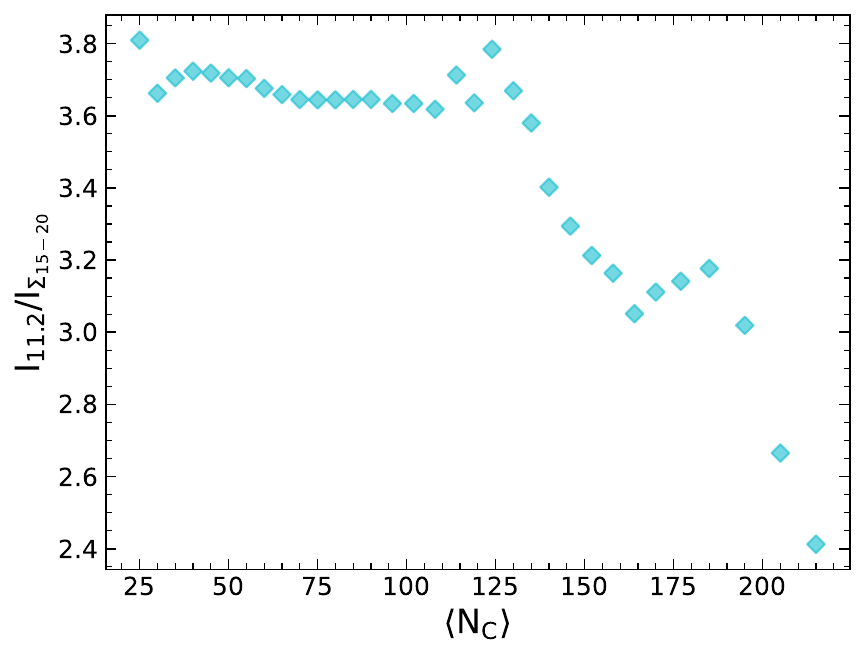} 
    \caption{The dependency of $I_{3.3}$ (blue), $I_{6.2}$ (orange), $I_{7.7}$ (green), $I_{8.6}$ (red), and $I_{11.2}$ (cyan) over $I_{\Sigma_{15-20}}$ on \AvgNc{}. Top left:  X/$I_{\Sigma_{15-20}}$ PAH intensity ratios vs \AvgNc; top middle: $I_{3.3}$/$I_{\Sigma_{15-20}}$ vs \AvgNc; Top right: $I_{6.2}$/$I_{\Sigma_{15-20}}$ vs \AvgNc; Bottom left: $I_{7.7}$/$I_{\Sigma_{15-20}}$ vs \AvgNc; Bottom middle: $I_{8.6}$/$I_{\Sigma_{15-20}}$ vs \AvgNc; Bottom right: $I_{11.2}$/$I_{\Sigma_{15-20}}$ vs \AvgNc. The discontinuities around $\sim$110 \AvgNc{} are due to the different data sets used, each corresponding to a different case of PAH processing (see Section~\ref{sec:method_analysis} and \citetalias{Maragkoudakis2023} for details).}
    \label{fig:X_F15_20}
\end{figure*}

The presence of the $I_{3.3}/X$ correlations is attributed to the 3.3~\micron{} PAH band intensity,  which has the strongest dependence on \Nc{} among the different bands, as demonstrated by \citetalias{Maragkoudakis2020}. The 3.3~\micron{} band involves the stretching vibration of the C-H bonds and has the least dependence on molecular structure, making it the main driver behind the correlations of the intensity ratio of the solo CH out-of-plane to CH stretching mode with \Nc{} for neutral, cationic and anionic PAHs. The strong dependence of the 3.3~\micron{} PAH band on \AvgNc{} is also presented in \citetalias{Maragkoudakis2023} for spectra of PAH mixtures following a size distribution, with the $I_{3.3}$/$I_{11.2}$ ratio presenting a monotonic decline and fully separated values across the \AvgNc{} range. Although the $I_{3.3}$/$I_{8.6}$ ratio shows a strong correlation with \AvgNc{}, in astronomical spectra the 8.6 \micron{} PAH feature can be much more severely impacted by the presence of silicate absorption at 9.7 \micron{}, and in combination with the band's proximity to the 7.7 \micron{} PAH complex, the decomposition and recovery of the feature can become challenging. In addition, the 6.2, 7.7, and 8.6 \micron{} PAH bands are produced from the contribution of both neutral and cationic PAHs, thus have a dependence on charge state, whereas the 3.3 and 11.2 \micron{} PAH bands are predominantly due to neutral PAHs. Therefore, the similar dependence of $I_{3.3}$ and $I_{11.2}$ on charge state, as well as the straightforward recovery and measurement of the respective intensities, makes the $I_{3.3}$/$I_{11.2}$ PAH band ratio an optimal PAH size tracer for the small-to-intermediate PAH population which contributes significantly to the 3-12~\micron{} PAH spectrum. 

\subsection{Alternative and qualitative PAH size tracers} \label{subsec:alt_tracers}

In cases where observations cover a wavelength range without the 3.3~\micron{} PAH band, alternative PAH size tracers are needed. Ideally, these tracers should have considerable wavelength separation to be most sensitive to the size. Smaller PAHs will attain a higher temperature, and thus emit stronger at shorter wavelengths, than larger PAHs upon the absorption of the same photon energy, because they have to redistribute the absorbed energy over a smaller number of available vibrational modes. The 15-20~\micron{} astronomical PAH emission complex is attributed to PAHs significantly larger in size than those held responsible for the emission between 3--12 \micron{} \citep[e.g.,][]{VanKerckhoven:00, Bauschlicher2009, Boersma2010} and thus is a useful candidate for an alternative to the 3.3~\micron{} PAH band. With the 15-20~\micron{} PAH complex being beyond the 10-15~\micron{} region implies that small PAHs contribute even less here to the emission \citep[][]{Allamandola1989, Schutte1993, Bauschlicher2009}.

Figure~\ref{fig:X_F15_20} presents the dependency of of $I_{3.3}$/$I_{\Sigma_{15-20}}$, $I_{6.2}$/$I_{\Sigma_{15-20}}$, $I_{7.7}$/$I_{\Sigma_{15-20}}$, $I_{8.6}$/$I_{\Sigma_{15-20}}$, and $I_{11.2}$/$I_{\Sigma_{15-20}}$ on \AvgNc. As expected, the $I_{3.3}$/$I_{\Sigma_{15-20}}$ ratio shows the most well-defined behaviour, spanning a large dynamic range. The $I_{8.6}$/$I_{\Sigma_{15-20}}$ and $I_{11.2}$/$I_{\Sigma_{15-20}}$ ratios show a weak and non-monotonic dependence on \AvgNc{}, due to the dependence of $I_{8.6}$ and $I_{11.2}$ on the number of adjacent solo hydrogens, which has a non-monotonic dependence on \AvgNc{} \citep[][]{Ricca2018}. Furthermore, for PAHs with straight edges (i.e. with adjacent solo hydrogens) that have 8.6 and 11.2 \micron{} bands, the introduction (or removal) of bay regions on the periphery, which decreases (or increases) the number of solo hydrogens, will have a higher impact on these bands than on the $I_{6.2}$ and $I_{7.7}$ bands.
The $I_{7.7}$/$I_{\Sigma_{15-20}}$ and $I_{6.2}$/$I_{\Sigma_{15-20}}$ intensity ratios are relatively well defined, mostly decreasing monotonically up to \AvgNc{} $\sim$164-170, which flatten, and then continue decreasing. Such fluctuations in the intensity ratios are due to the chemical diversity within a bin, i.e., the relative number of PAHs with straight edges, non-compact PAHs with eroded edges, and elongated PAHs. In addition, \cite{Ricca2012} note that for PAHs larger than 150 carbons the profiles of the 6.2, 7.7, and 8.6~\micron{} bands show significant spectral variations compared to those in astronomical observations, indicating an upper limit to the size of astronomical PAHs.

The nearly monotonic dependence of $I_{6.2}$/$I_{\Sigma_{15-20}}$ and $I_{7.7}$/$I_{\Sigma_{15-20}}$  on \AvgNc{} can be of value for observations that do not cover the 3.3~\micron{} region, such as \textit{JWST} MIRI(-MRS)-only observations, assuming sources with comparable PAH size distributions with the described models, i.e., for small-to-intermediate sized PAH populations or for the lower-to-intermediate part of their size distribution (see Section~\ref{subsec:observations}). For both optimal and alternative PAH size proxies, determination of the PAH ionisation fraction is required, typically estimated using proxies such as the $I_{7.7}$/$I_{11.2}$ (e.g., \citealt{Allamandola1999}; \citealt{Galliano2008a}; \citetalias{Maragkoudakis2020}; \citetalias{Maragkoudakis2023}).

\subsection{Application to observations} \label{subsec:observations}

We applied the current models to literature work that combines AKARI and \textit{Spitzer}/IRS observations, thus includes the 3.3~\micron{} PAH feature from AKARI spectra along with the PAH bands in the 5-20~\micron{} region from \textit{Spitzer}/IRS. Two independent data sets were employed: (i) 133 galaxies from the GOALS sample \citep[][]{Armus2009, Stierwalt2014, Inami2018}, and (ii) 113 galaxies from \cite{Lai2020}. The GOALS sample is comprised of LIRGs and ULIRGs spanning a range in merger stage and morphology, while the \cite{Lai2020} sample consists mostly of star-forming galaxies. For the GOALS sample, fixed aperture widths were used to extract AKARI 1D spectra, comparable to the SL \textit{Spitzer}/IRS aperture size. The extinction corrected 3.3~\micron{} PAH bands were collected from \cite{Inami2018}, which were fitted with a Gaussian and a linear function for the local continuum. For the 3.3~\micron{} PAH intensities we further applied the scaling factor described in \cite{McKinney2021} to allow comparison with the \textit{Spitzer}/IRS PAH features from \cite{Stierwalt2014}, which were modelled using the CAFE spectral decomposition code \citep{Marshall2007}. \cite{Lai2020} scaled the AKARI and \textit{Spitzer}/IRS spectra to match the WISE photometry in order to stitch the separate spectral segments, and modelled the PAH band features using the PAHFIT spectral decomposition code \citep{Smith07b}. 

Accurate determination of PAH size requires information on the PAH ionisation fraction within a source, as the different PAH bands have a varying response to ionisation. For the GOALS sample, \cite{McKinney2021} employed the PAH charge-size grid of \citetalias{Maragkoudakis2020} for a 10~eV photon energy, and demonstrated that the galaxies allocate the region between the N25C75 and purely cationic PAHs. \cite{Lai2020} provide information only for the 6.2~\micron{} PAH band, and thus direct application of the \citetalias{Maragkoudakis2020} grid, which uses the $I_{11.2}$/$I_{7.7}$ PAH band intensity ratio for the PAH ionisation fraction, could not be done. Instead, we used the parameters defining the correlation between $I_{6.2}$/$I_{11.2}$ and $I_{7.7}$/$I_{11.2}$ from \cite{Maragkoudakis2018b} to estimate $I_{7.7}$/$I_{11.2}$, and assuming an ISRF model grid we found that the galaxies also allocate the region between the N25C75 and purely cationic PAHs, similar to the GOALS sample.

\begin{figure*}
    \centering
    \includegraphics[scale=0.45]{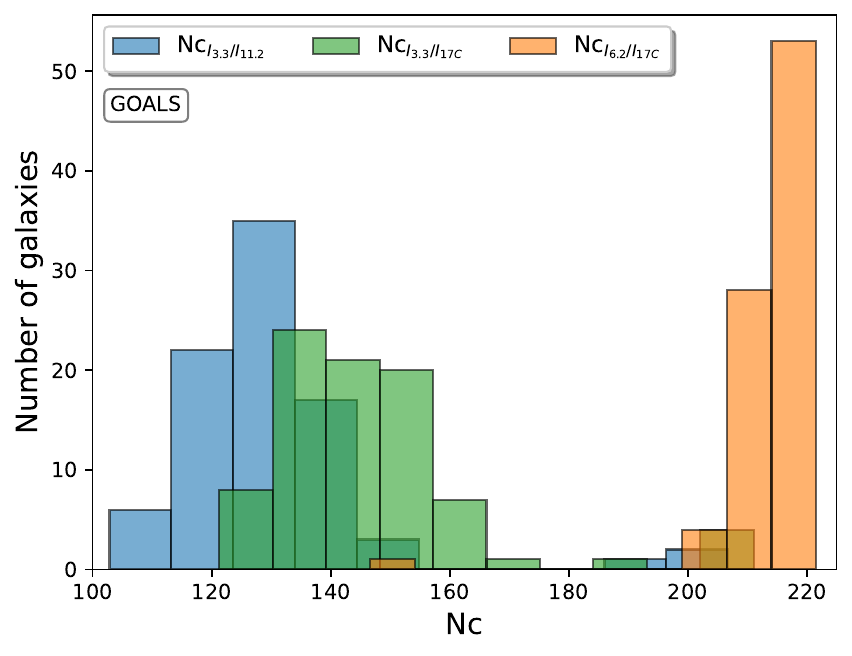}
    \includegraphics[scale=0.45]{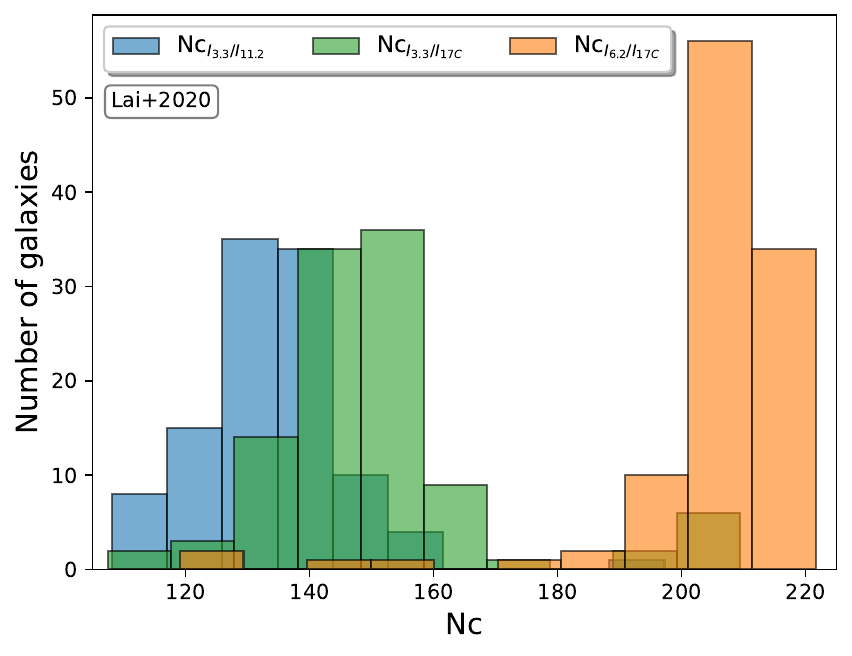}
    \caption{The \AvgNc{} distributions for the GOALS \protect\citep[][left panel]{Stierwalt2014, Inami2018}{} and \protect\citeauthor{Lai2020} (\citeyear{Lai2020}, right panel) samples, recovered from the $I_{3.3}$/$I_{11.2}$ (blue colour), $I_{3.3}$/$I_{\Sigma_{15-20}}$ (green colour), and $I_{6.2}$/$I_{\Sigma_{15-20}}$ (orange colour) --vs \AvgNc{} calibrations.}
    \label{fig:histograms}
\end{figure*}

Figure \ref{fig:histograms} presents the \AvgNc{} distributions in the two samples, recovered when employing the $I_{3.3}$/$I_{11.2}$, $I_{3.3}$/$I_{\Sigma_{15-20}}$, and $I_{6.2}$/$I_{\Sigma_{15-20}}$ vs \AvgNc{} correlations for the N25C75 case (Figures~\ref{fig:F33_X_various_if} and \ref{fig:X_F15_20_various_if}), where each PAH band intensity ratio vs \AvgNc{} correlation was modelled with a 3rd-degree polynomial. For both samples, the PAH band intensity of the 17~\micron{} complex was provided and used instead of the $I_{\Sigma_{15-20}}$. The offset between the respective \AvgNc{} distributions, in both samples, indicates variation between the PAH size distribution among the models and the galaxies. Thus, in this case the $I_{3.3}$/$I_{11.2}$ PAH band ratio provides a good description on the small-to-intermediate PAH population, while the $I_{6.2}$/$I_{\Sigma_{15-20}}$ ratio provides insight into the higher end of the PAH size distribution within these sources, as expected \citep[e.g.,][]{Schutte1993}. Such a comparison, between the \AvgNc{} distributions recovered using the current models which are sensitive to the small-to-intermediate PAH populations, can be used to assess the presence (or absence) of a large PAH population in a source.  

Interestingly, the offset between the respective \AvgNc{} distributions is similar for both samples, which opens the possibility for cross-calibration between the size tracers. Specifically, the $\langle\mathrm{N_{C}}\rangle_{I_{3.3}/I_{11.2}}$ - $\langle\mathrm{N_{C}}\rangle_{I_{3.3}/I_{\Sigma_{15-20}}}$ offset is on average $\sim 15$ \AvgNc, and the $\langle\mathrm{N_{C}}\rangle_{I_{3.3}/I_{11.2}}$ - $\langle\mathrm{N_{C}}\rangle_{I_{6.2}/I_{\Sigma_{15-20}}}$ offset is on average $\sim 80$ \AvgNc. Despite differences among the physical properties of the 246 galaxies in both samples combined, observations covering the $I_{6.2}$ and $I_{\Sigma_{15-20}}$ PAH features can provide information on the size of the small-to-intermediate PAH population employing the presented model, and offset by ${\sim}80$ \AvgNc. We note that this average offset appears to be insensitive to the applied decomposition method. 

In addition, although the $I_{3.3}/$X vs \AvgNc{} correlations may predict higher absolute \Nc{} values when applied to observations\footnote{Due to lack of proper treatment of anharmonicity effects, affecting mostly the calculation of the intrinsic strength of the C-H modes in neutral PAHs \citep{Mackie2022}}, both the $I_{3.3}/$X and X$/I_{\Sigma_{15-20}}$ correlations can provide insights into the relative difference and variance of \AvgNc{} between e.g., different regions within a source, or between similar type of sources, independently of delivering an absolute \AvgNc{} value. 

Finally, a detailed cross-calibration between the different PAH size proxies requires homogeneous and high quality observations, such as those delivered by the \textit{JWST}, covering all the key PAH features, and for consistency, following a uniform reduction and spectral decomposition method. In addition, the ongoing expansion of PAHdb's library to include larger PAH species as well as PAH clusters, will aid to the expansion and refinement of the current models and the related PAH size proxies.    

\section{Summary and Conclusions} \label{sec:conclusions}

We examined the dependency of PAH band intensity ratios on \AvgNc{} using spectra synthesised from mixtures of PAHs that follow a size distribution and at different ionisation fractions, and assessed their effectiveness as tracers for PAH size. Our results and conclusions are:

\renewcommand{\theenumi}{\textit{\roman{enumi})}}%
\begin{enumerate}

\item PAH band intensity ratios that include the 3.3~\micron{} PAH band are optimal PAH size tracers, following the efficiency of the $I_{3.3}$/$I_{11.2}$ ratio as demonstrated in \citetalias{Maragkoudakis2023}, and show a monotonic dependence on \AvgNc{} consistent with the single-PAH molecule behaviour presented in \citetalias{Maragkoudakis2020}. These ratios traces the small-to-intermediate sized PAHs, with the $I_{3.3}$/$I_{11.2}$ ratio considered the better choice given the similar dependence of the $I_{3.3}$ and $I_{11.2}$ PAH bands on charge state. \\

\item Band intensity ratios that include features between 15-20~\micron{} and those with large wavelength separation, compared to 15-20~\micron{}, i.e., the 3.3, 6.2, and 7.7~\micron{} bands, trace PAH size and can serve as alternative tracers for the small-to-intermediate sized PAHs, either directly assuming sources with a PAH size distribution on par with the given models, or through application of a cross-calibration factor. \\

\item Using PAH band measurements of 246 galaxies from two separate samples, we derived a cross-calibration between the $I_{3.3}$/$I_{11.2}$ and $I_{6.2}$/$I_{\Sigma_{15-20}}$ PAH size tracers, for the estimation of \AvgNc{} of small-to-intermediate PAH populations, for sources with a different PAH size distribution compared to the presented model. \\

\item The dependence of the $I_{6.2}$/$I_{\Sigma_{15-20}}$ and $I_{7.7}$/$I_{\Sigma_{15-20}}$ intensity ratios on \AvgNc{} allows characterisation of the PAH size and the PAH size distribution for small-to-intermediate sized PAHs, without the need for 3.3~\micron{} coverage, for example, with \textit{JWST} MIRI(-MRS)-only observations, for sources with comparable PAH size distributions with the described models, i.e., for small-to--intermediate PAH populations, or for the lower-to-intermediate part of their size distribution.  

\end{enumerate}

\section*{Acknowledgements}

We thank the anonymous referee for the constructive comments that have improved the clarity of the paper. A.M.'s research was supported by an appointment at NASA Ames Research Center, administered by the Bay Area Environmental Research Institute, and an appointment to the NASA Postdoctoral Program at NASA Ames Research Center, administered by the Oak Ridge Associated Universities through a contract with NASA.  E.P. acknowledges support from an NSERC Discovery Grant and a Western Science and Engineering Research Board (SERB) Accelerator Award. C.B. is grateful for an appointment at NASA Ames Research Center through the San Jos\'e State University Research Foundation (80NSSC22M0107). A.M., A.R. and C.B. gratefully acknowledge support from the ``NASA Ames Laboratory Astrophysics Directed Work Package (LADWP) Round 2 ISFM''.

\section*{Data Availability}

The analysis products of this work will be shared on a reasonable request to the corresponding author.



\bibliographystyle{mnras}
\input{Bibliography.bbl} 




\appendix

\section{PAH Size Tracers for Different Ionisation Fractions} \label{app:var_fi}

\begin{figure*}
    \centering
    \includegraphics[scale=0.45]{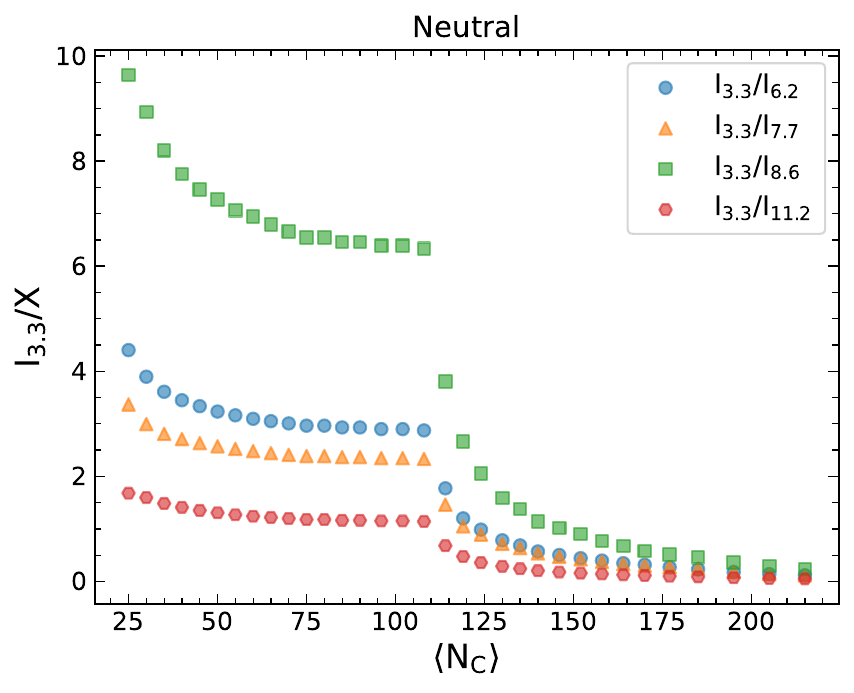}
    \includegraphics[scale=0.45]{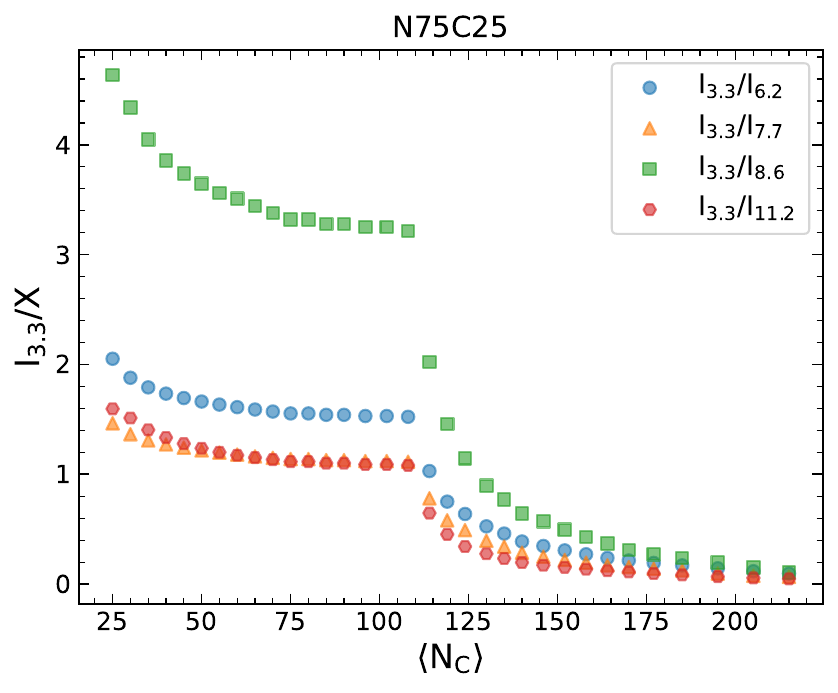} \\
    \includegraphics[scale=0.45]{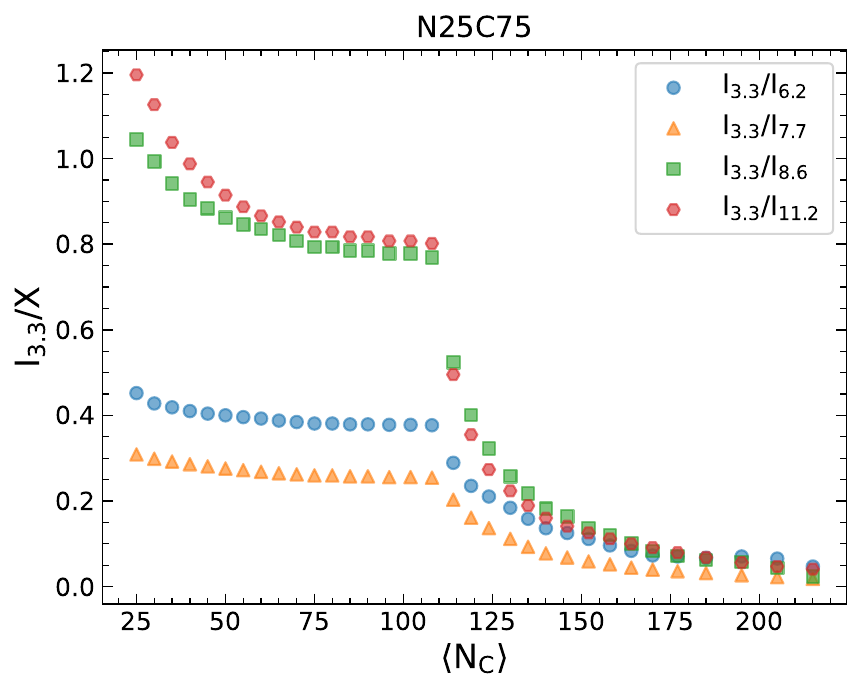} 
    \includegraphics[scale=0.45]{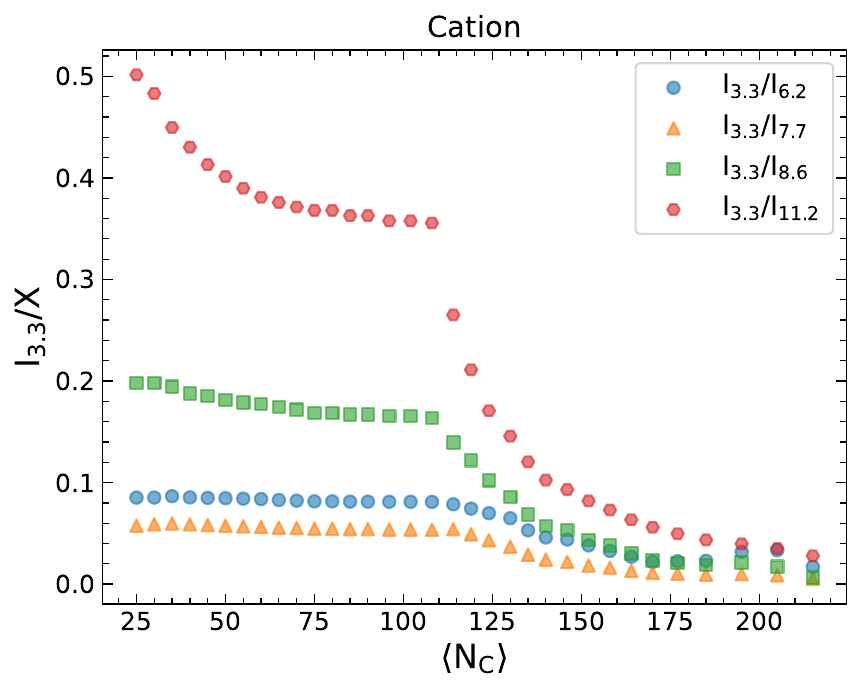} 
    \caption{The dependence of $I_{3.3}$ over $I_{6.2}$ (blue), $I_{7.7}$ (orange), $I_{8.6}$ (green), and $I_{11.2}$ (red) PAH band intensities on \AvgNc{}, for different ionisation fractions. The discontinuities at $\sim$110 \AvgNc{} are due to the different data sets used, each corresponding to a different case of PAH processing (see Section~\ref{sec:method_analysis} and \citetalias{Maragkoudakis2023} for details).}
    \label{fig:F33_X_various_if}
\end{figure*}

\begin{figure*}
    \centering
    \includegraphics[scale=0.4]{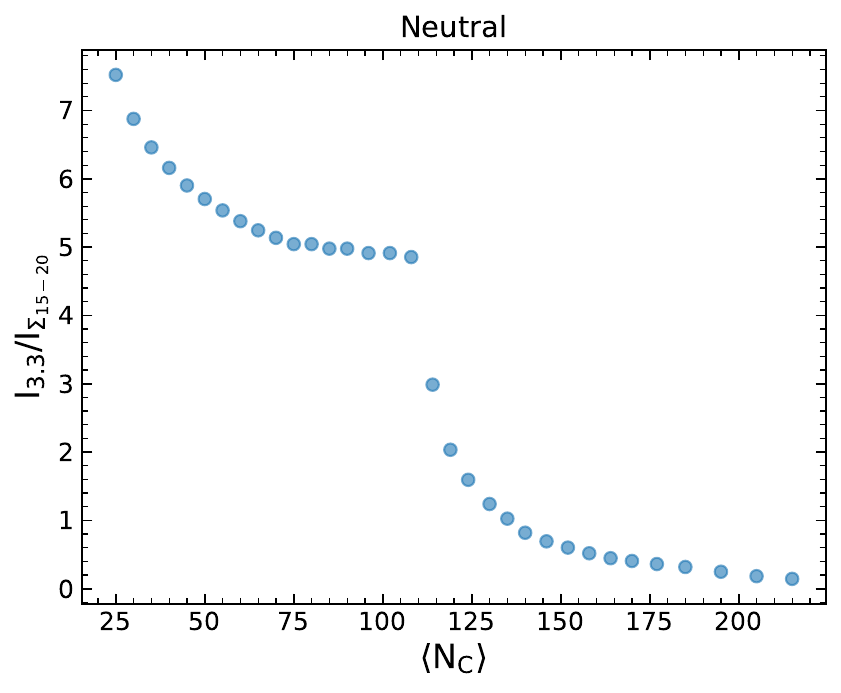}
    \includegraphics[scale=0.4]{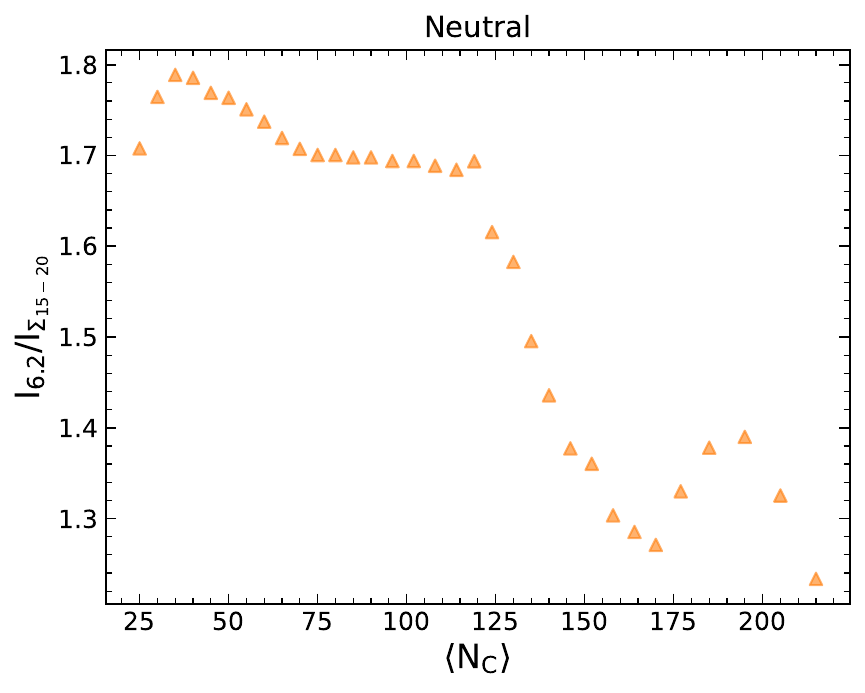}
    \includegraphics[scale=0.4]{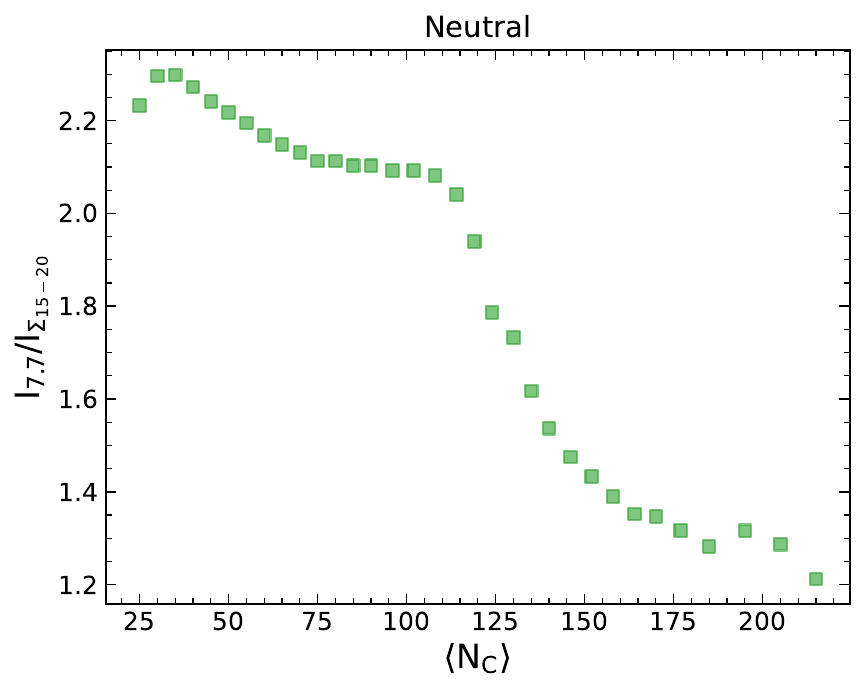} \\
    \includegraphics[scale=0.4]{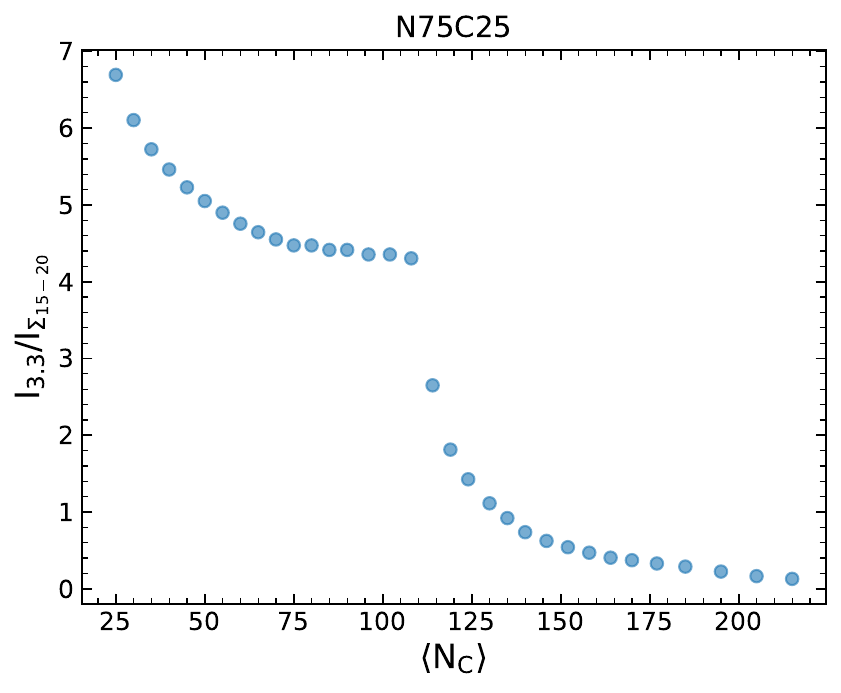}
    \includegraphics[scale=0.4]{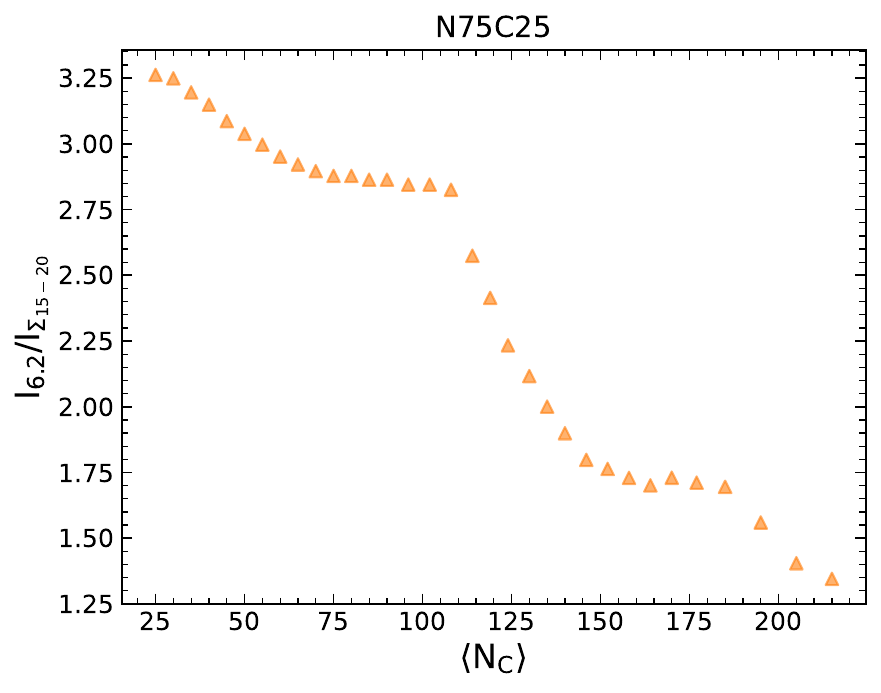}
    \includegraphics[scale=0.4]{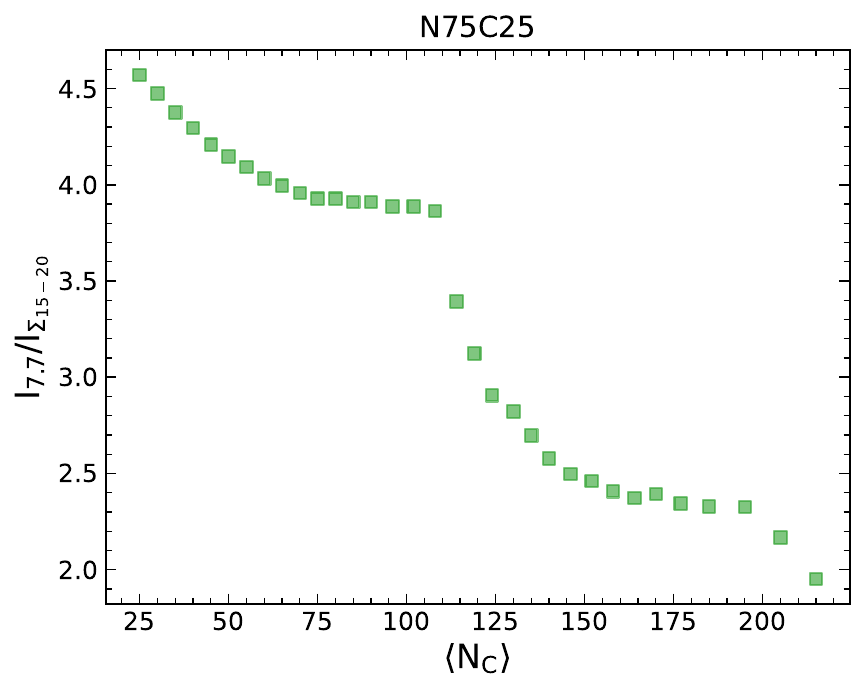} \\
    \includegraphics[scale=0.4]{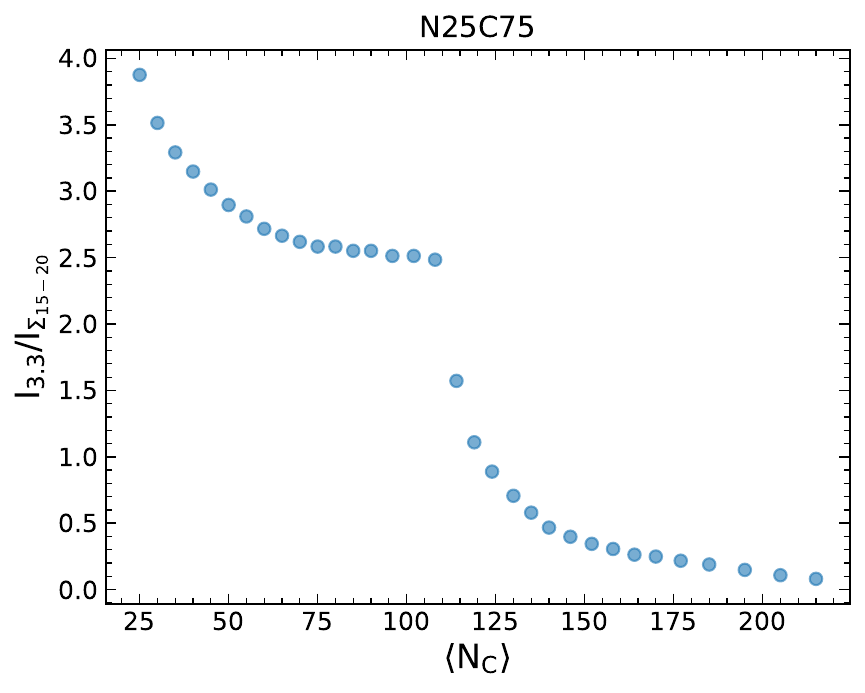}
    \includegraphics[scale=0.4]{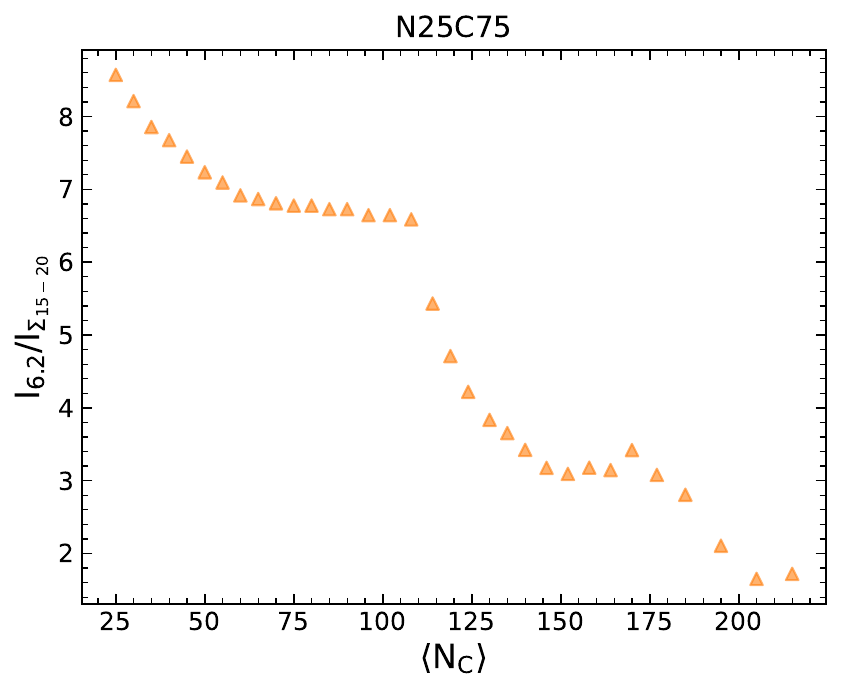}
    \includegraphics[scale=0.4]{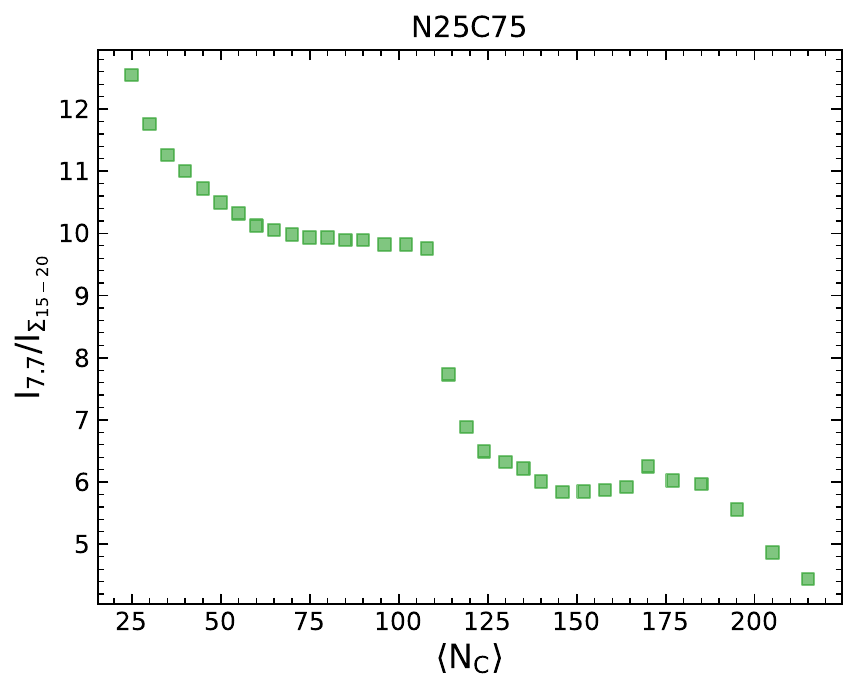} \\
    \includegraphics[scale=0.4]{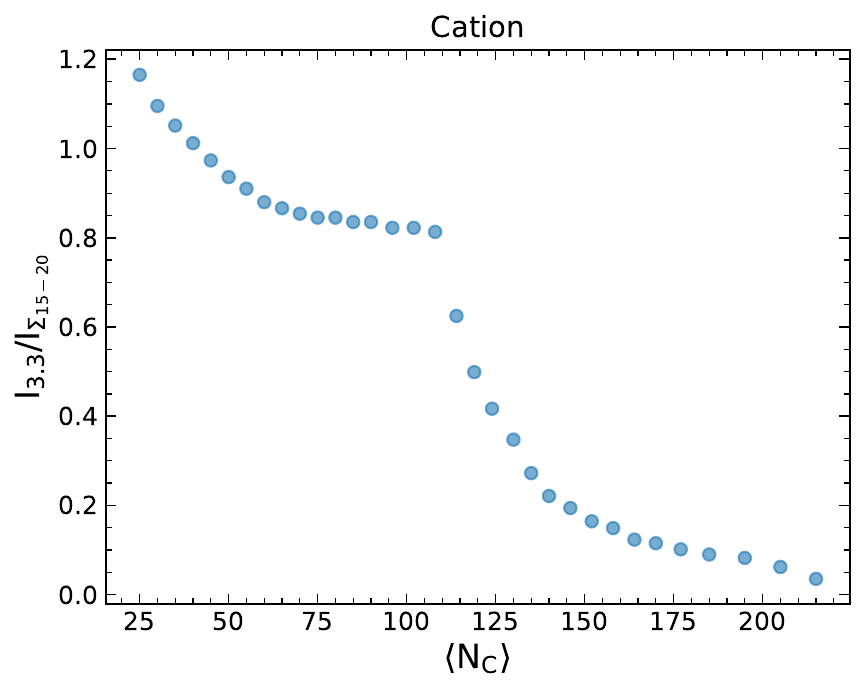}
    \includegraphics[scale=0.4]{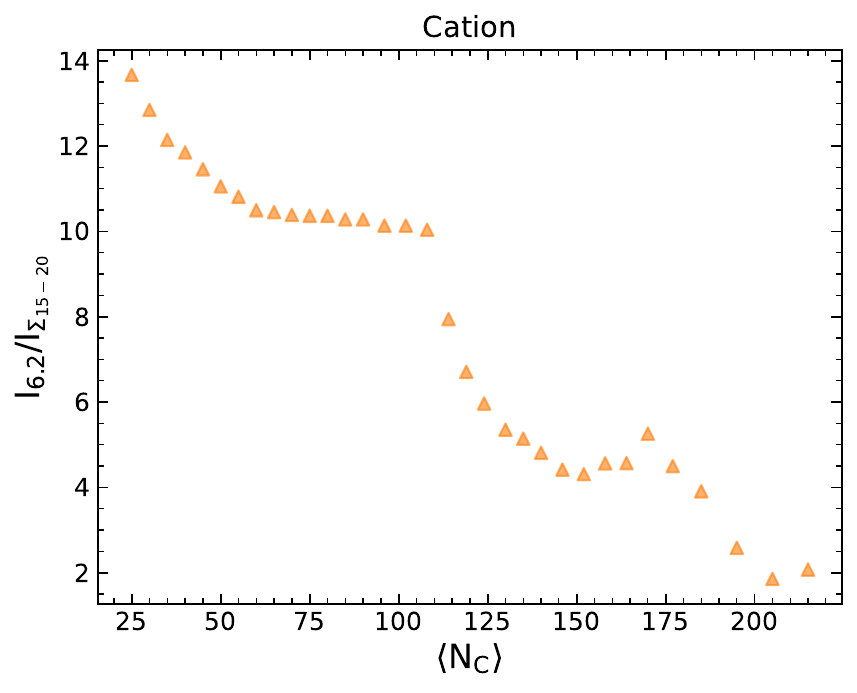}
    \includegraphics[scale=0.4]{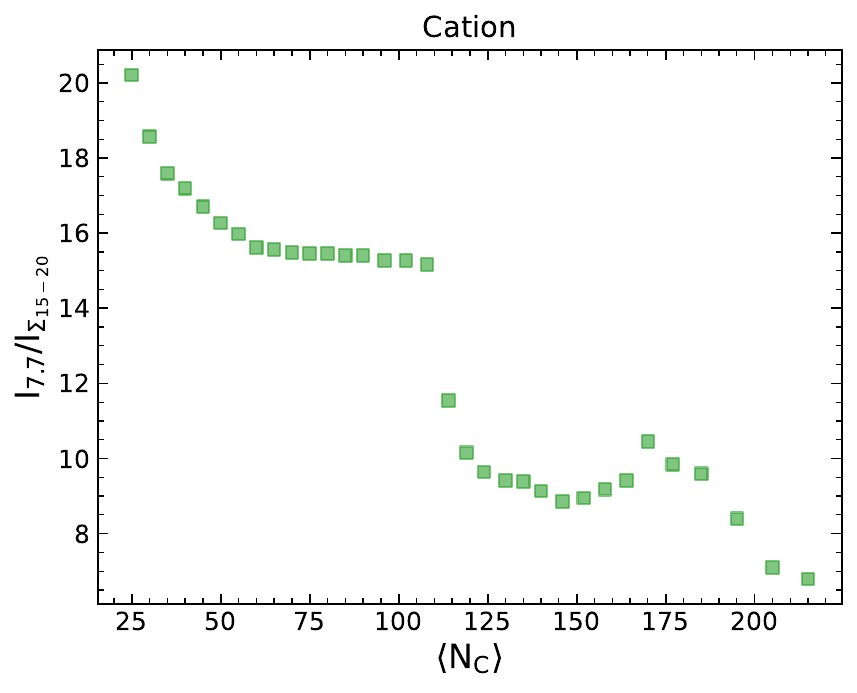}
    \caption{The dependency of $I_{3.3}$ (left panels), $I_{6.2}$ (middle panels), and $I_{7.7}$ (right panels), over $I_{\Sigma_{15-20}}$ on \AvgNc{}, for different ionisation fractions. Top row: Neutral; Second row from the top: N75C25; Third row from the top: N25C75; Bottom row: Cation. The discontinuities at $\sim$110 \AvgNc{} are due to the different data sets used, each corresponding to a different case  of PAH processing (see Section~\ref{sec:method_analysis} and \citetalias{Maragkoudakis2023} for details).}
    \label{fig:X_F15_20_various_if}
\end{figure*}

The dependence of the various PAH intensity ratios involving the $I_{3.3}$ and $I_{\Sigma_{15-20}}$ intensities, as described in Sections~\ref{subsec:optimal_tracers} and \ref{subsec:alt_tracers}, are presented in Figures~\ref{fig:F33_X_various_if} and \ref{fig:X_F15_20_various_if}, respectively, for different ionisation fractions, i.e., purely neutral, 75-25 percent neutral-cation (N75C25), 25-75 percent neutral-cation (N25C75), and purely cationic PAHs. The $I_{3.3}/X$ ratios have the most uniform and monotonic dependence on \AvgNc, similar to the N50C50 case. The $I_{6.2}/I_{\Sigma_{15-20}}$ and $I_{7.7}/I_{\Sigma_{15-20}}$ intensity ratios are mostly monotonic, with fluctuations present in the \AvgNc{} $\sim$160-200 range that become more apparent when moving progressively to fully cationic PAH populations. For the pure neutral PAHs the $I_{6.2}/I_{\Sigma_{15-20}}$ ratio is unreliable and the $I_{7.7}/I_{\Sigma_{15-20}}$ ratio offers a better description among the two ratios. 

\bsp	
\label{lastpage}
\end{document}